\documentclass[prb,showpacs,floatfix,twocolumn]{revtex4}
\usepackage{graphicx}

\begin{document}

\title{Magnetostructural transition, metamagnetism, and magnetic phase coexistence in Co$_{10}$Ge$_3$O$_{16}$}
%beneath $T$ = 20\,K
%Magnetostructural and field-induced phase transitions in Co$_{10}$Ge$_3$O$_{16}$
%Magnetostructural transitions and metamagnetism induced by Ising spins in the spinel-rock salt intergrowth Co$_{10}$Ge$_3$O$_{16}$

\author{Phillip T. Barton} \email{pbarton@mrl.ucsb.edu}
\author{Ram Seshadri}
\affiliation{Materials Department and Materials Research Laboratory, University of California, Santa Barbara, CA, 93106, USA}

\author{Anna Llobet}
\affiliation{Lujan Neutron Scattering Center, Los Alamos National Laboratory, Mail Stop H805, Los Alamos, NM 87545, USA}

\author{Matthew R. Suchomel}
\affiliation{X-Ray Science Division, Argonne National Laboratory, Argonne, IL 60439, USA}

\date{\today}

\begin{abstract}

Co$_{10}$Ge$_3$O$_{16}$ crystallizes in an intergrowth structure featuring alternating layers of spinel and rock salt. Variable-temperature powder synchrotron X-ray and neutron diffraction, magnetometry, and heat capacity experiments reveal a magnetostructural transition at $T_{\rm{N}}$ = 203\,K. This rhombohedral-to-monoclinic transition involves a slight elongation of the CoO$_6$ octahedra along the apical axis. Below $T_{\rm{N}}$, the application of a large magnetic field causes a reorientation of the Co$^{2+}$ Ising spins. This metamagnetic transition is first-order as evidenced by a latent heat observed in temperature-dependent measurements. This transition is initially seen at $T$ = 180\,K as a broad upturn in the $M$-$H$ near $H_{\rm{C}}$ = 3.9\,T. The upturn sharpens into a kink at $T$ = 120\,K and a ``butterfly'' shape emerges, with the transition causing hysteresis at high fields while linear and reversible behavior persists at low fields. $H_{\rm{C}}$ decreases as temperature is lowered and the loops at positive and negative fields merge beneath $T$ = 20\,K. The antiferromagnetism is described by $k_{\rm{M}}$ = (00$\frac{1}{2}$) and below $T$ = 20\,K a small uncompensated component with $k_{\rm{M}}$ = (000) spontaneously emerges. Despite the Curie-Weiss analysis and ionic radius indicating the Co$^{2+}$ is in its high-spin state, the low-temperature $M$-$H$ trends toward saturation at $M_{\rm{S}}$ = 1.0 $\mu_{\rm{B}}$/Co. We conclude that the field-induced state is a ferrimagnet, rather than a $S$ = 1/2 ferromagnet. The unusual $H$-$T$ phase diagram is discussed with reference to other metamagnets and Co(II) systems.
%Curie-Weiss analysis suggests that the Co$^{2+}$, with $S$ = 3/2 and $L$ = 3, acts as a Kramer's doublet due to spin-orbit coupling.
\pacs{
	75.30.Kz %Magnetic phase boundaries (including magnetic transitions, metamagnetism, etc.)
	75.50.Ee %Antiferromagnetic materials
	75.60.Nt %Magnetic annealing and temperature-hysteresis effects
	75.50.Gg %Ferrimagnetics
}
\end{abstract}
\maketitle

\section{Introduction} 

Compounds featuring Co$^{2+}$ often display interesting magnetic field-induced phase transitions.\cite{Raveau_Wiley} The driving force behind these transitions is the Ising-like single-ion anisotropy of octahedral Co$^{2+}$ that results from spin-orbit-lattice coupling. Co$^{2+}$ is $d^7$ and in an octahedral coordination environment it typically adopts a high-spin $^4$F$_{9/2}$ electronic ground state ($S$ = 3/2 and $L$ = 3); with spin-orbit coupling it is often described as a Kramer's doublet with $J_{\rm{eff}}$ = 1/2. The complex magnetism that results from this electronic state is seen in many types of Co(II)-containing antiferromagnets including the prototypical CoCl$_{2}\cdot$2H$_2$O,\cite{Kobayashi_JPSJ64} $\beta$-Co(OH)$_2$,\cite{Neilson_PRB11} hybrid metal-organic frameworks,\cite{Kurmoo_CSR09}, and oxyselenides.\cite{Melot_JPCM10} Co(II) is also a popular species to study in the field of molecular magnets where it is of interest for spin-crossover or spin-state transitions.\cite{Gatteschi_AM94} The majority of the aforementioned anisotropic metamagnets exhibit spin reorientations and their phase diagrams are well described by an Ising model with magnetic exchange that is antiferromagnetic for nearest neighbors and ferromagnetic for next-nearest neighbors.\cite{Stryjewski_AP77,Carlin_ACR80} These anisotropic systems are distinct from those with Heisenberg spins which undergo spin-flop transitions where spin canting is important. Field-induced transitions also occur in systems with ferrimagnetic or noncollinear ground states such as CoCr$_2$O$_4$.\cite{Tsurkan_PRL13} These ground states are not as well investigated as N\'{e}el antiferromagnetism, but are more technologically interesting. Some of these systems have complex $H$-$T$ phase diagrams that have been explored by experimentalists and are begininning to be explained by theoriticians. In this manuscript, we investigate the complex behavior of the relatively unstudied Co$_{10}$Ge$_3$O$_{16}$.

%demagnetizing effects
%tri-critical point where phase transition goes from first to second order

%A good discussion of field-dependent magnetic phenomena, also termed metamagnetism and/or spin flip/flop is given by Carlin and van Duyneveldt.\cite{Carlin_ACR80} Stryjewski and Giordano provide a thorough review of anisotropic magnetism, however their presented phase diagrams differ from that observed here.

%Compare to CoGeO3

\begin{figure}
\centering
\includegraphics[width=2.5in]{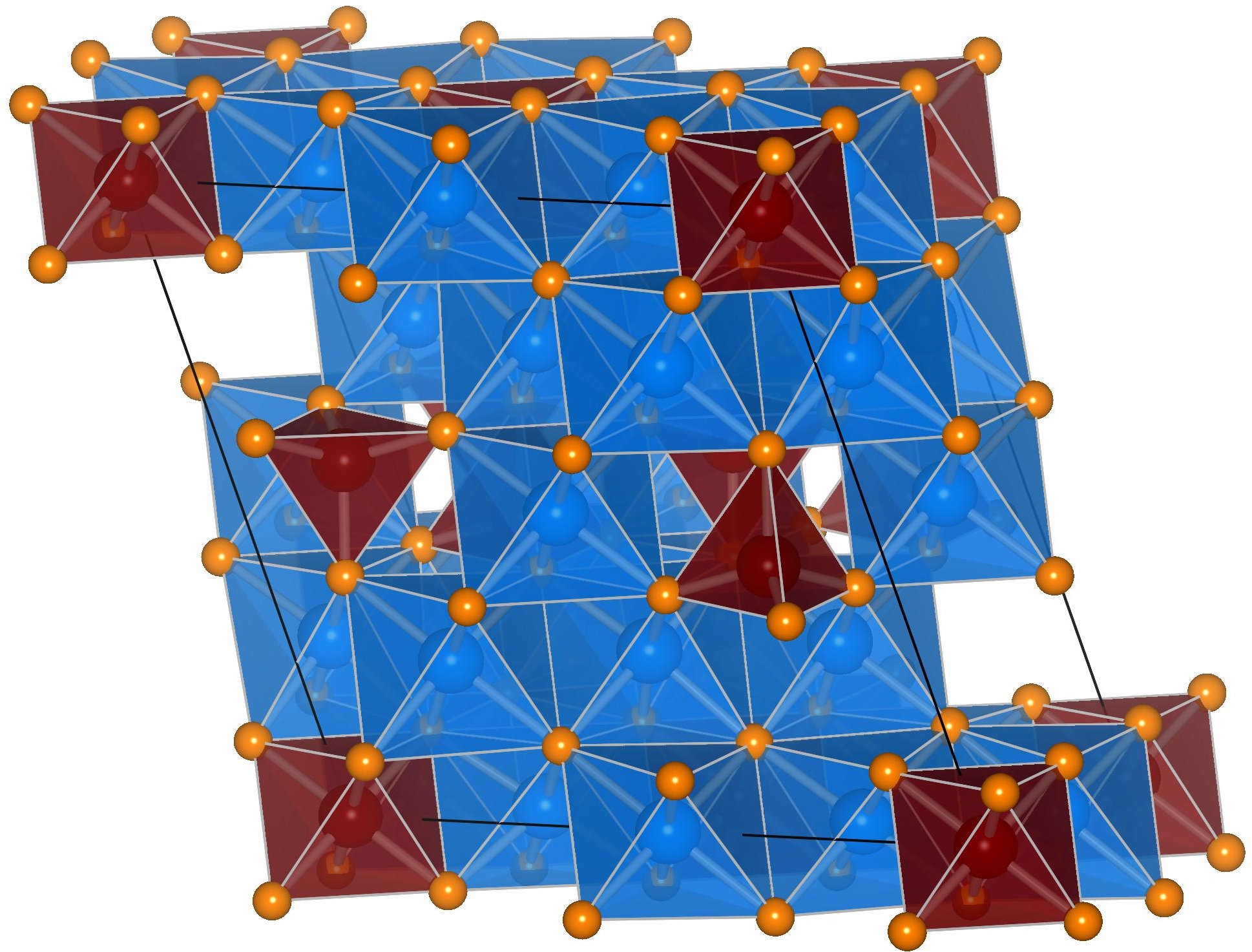}\\
\caption{(Color online) Schematic crystal structure of Co$_{10}$Ge$_3$O$_{16}$. This is the low-temperature $C$2/$m$ structure which has the same topology as the high-temperature $R\bar{3}m$, but a smaller unit cell. The spinel layer, with tetrahedral Ge$^{4+}$, is in the middle and is sandwiched between rock-salt-like slabs of edge-sharing Ge$^{4+}$ and Co$^{2+}$ octahedra. The sphere colors correspond to: maroon, Ge; blue, Co; and orange O.}
\label{fig:structure}
\end{figure}

Co$_{10}$Ge$_3$O$_{16}$ is an octahedral Co$^{2+}$ compound that is related to the spinel GeCo$_2$O$_4$ in both chemical composition and crystal structure. It was first prepared by Barbier who solved its structure from single-crystal X-ray diffraction.\cite{Barbier_AC95} Its layered crystal structure is an intergrowth composed of alternating spinel units and edge-sharing octahedra layers (Figure~\ref{fig:structure}). The O$^{2-}$ anions are close-packed, with a nearly ideal $c$/$a$ ratio, and the Ge$^{4+}$ sit in both tetrahedral and octahedral sites. Barbier's study found that bond valence sums indicate the expected ionic states and that the thermal parameters are almost isotropic. This structure type is related to the cation-deficient mineral aerugite, Ni$_{8.5}$As$_3$O$_{16}$.\cite{Davis_MM65,Fleet_AC89} Udod and co-workers concluded that Co$_{10}$Ge$_3$O$_{16}$ is a ferrimagnet based on magnetometry measurements,\cite{Udod_PSS07} however our study indicates that the exceedingly complex magnetism is not captured by the term ``ferrimagnetism''.

%Some Co(II) compounds develop sizable hysteresis loops at low temperature, but some don't - why?

Here we present physical property measurements including, magnetometry, heat capacity, and powder neutron diffraction, that indicate an antiferromagnetic transition in Co$_{10}$Ge$_3$O$_{16}$ at $T_N$ = 203\,K. A structural transition from rhombohedral to monoclinic symmetry accompanies the onset of magnetic order, as seen in powder synchrotron X-ray diffraction. A magnetic-field-induced phase transition occurs below $T_N$ and gives rise to hysteresis in both the $M$-$H$ and $\chi$-$T$ sweeps, indicating a first-order nature. Powder neutron diffraction elucidates the low-temperature magnetism by revealing the spontaneous emergence, below $T$ = 20\,K, of a small ferri- or ferromagnetic component with $k_M$ = (000) that coexists with the antiferromagnetism. We discuss the ground state of Co$_{10}$Ge$_3$O$_{16}$ with respect to the related spinel GeCo$_2$O$_4$ and other metamagnetic Co(II) systems.

\begin{figure}
\centering
\includegraphics[width=0.7\columnwidth]{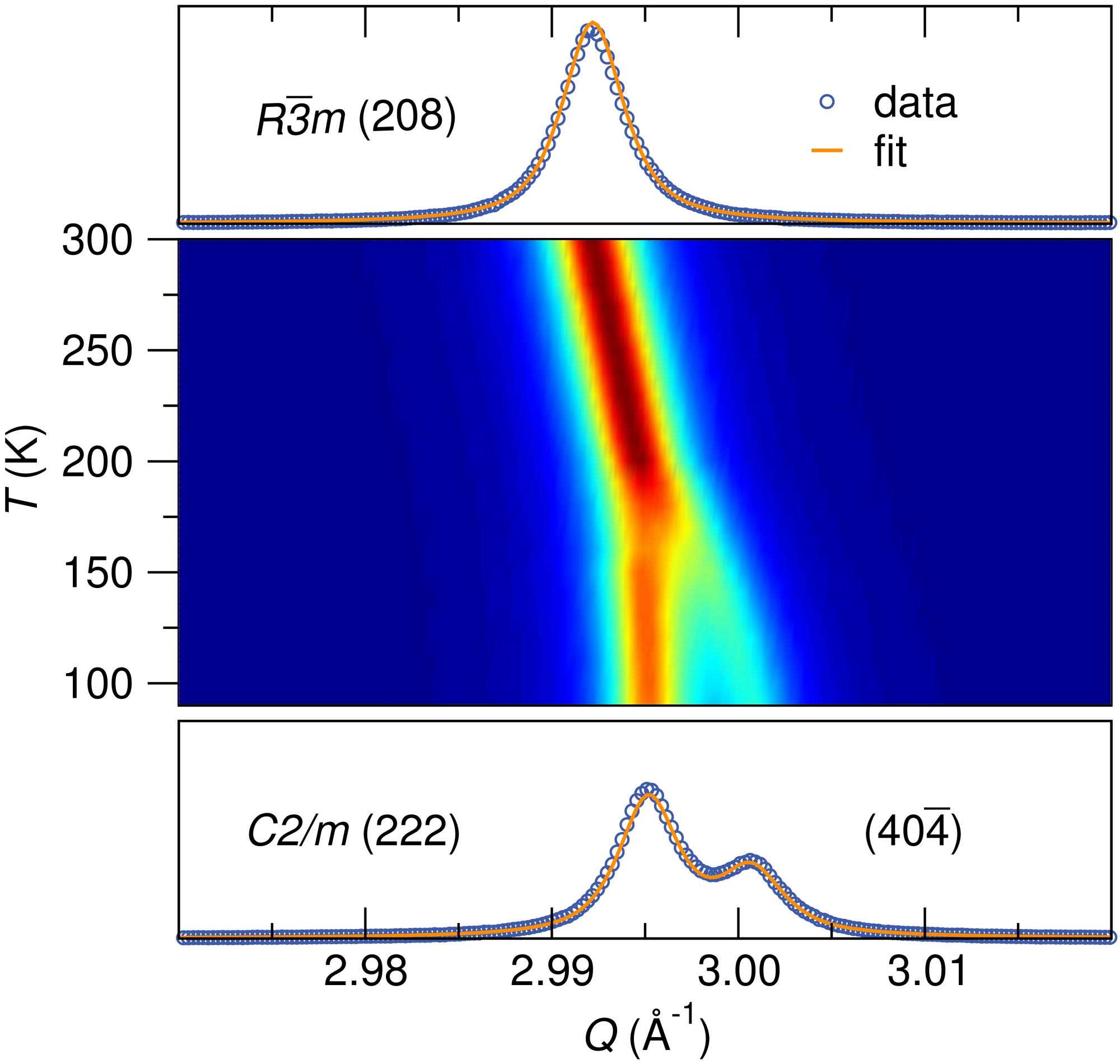}\\
\caption{(Color online) Temperature evolution of a powder synchrotron X-ray diffraction peak through the magnetostructural transition of Co$_{10}$Ge$_3$O$_{16}$. The top panel shows the $T$ = 300\,K data and fit for the $R\bar{3}m$ (208) peak. An intensity contour plot is displayed in the center panel to illustrate the splitting of the peak as temperature is lowered through $T_{\rm{N}}$ = 203\,K. The bottom panel presents the $T$ = 90\, K data and fit for the $C$2/$m$ (222) and (40$\bar{4}$) peaks.}
\label{fig:peak_evolution}
\end{figure}

\section{Methods}

Polycrystalline pellets of Co$_{10}$Ge$_3$O$_{16}$ were prepared by solid-state reaction of powder reagents at high temperature. Stoichiometric amounts of GeO$_2$ and Co$_3$O$_4$ were ground with an agate mortar and pestle, pressed at 100\,MPa, and fired in air at 1000$^{\circ}$C for 48\,h with an intermediate grinding. The pellets were placed on beds of powders of the same composition to avoid contamination by the ZrO$_2$ crucible. Powder synchrotron X-ray diffraction was performed at the 11-BM beamline ($\lambda$ = 0.413104 \AA) of the Advanced Photon Source, Argonne National Laboratory. Powder neutron diffraction was conducted at the HIPD beamline of the Lujan Neutron Scattering Center, Los Alamos National Laboratory. Rietveld\cite{Rietveld_JAC69} refinements were performed using GSAS/EXPGUI.\cite{Toby_JAC01} DICVOL, as implemented in FullProf, was used to index the low-temperature unit cell.\cite{Boultif_JAC04} ISODISTORT was used to explore the possible crystal distortion modes and to transform the unit cell atom positions to lower symmetry.\cite{Campbell_JAC06} Crystal structures were visualized using VESTA.\cite{Momma_VESTA_08} Magnetic properties were measured using a Quantum Design MPMS 5XL SQUID magnetometer and a PPMS DynaCool VSM. Heat capacity was measured using a thermal relaxation method as implemented in a Quantum Design PPMS. Electrical resistivity was measured by a PPMS DynaCool ETO to be 5$\times$10$^6$ $\Omega\cdot$cm at room-temperature and increased exponentially as temperature decreased, consistent with the expected insulating behavior of a brown-colored material.

%High-field magnetometry measurements up to 68\,T were conducted at the Los Alamos Pulsed Field Program of the National High Magnetic Field Lab.
%A 1% GeCo2O4 impurity was observed

\section{Results and Discussion}

\subsection{Crystal structure}

\begin{figure*}
\centering
\includegraphics[width=6in]{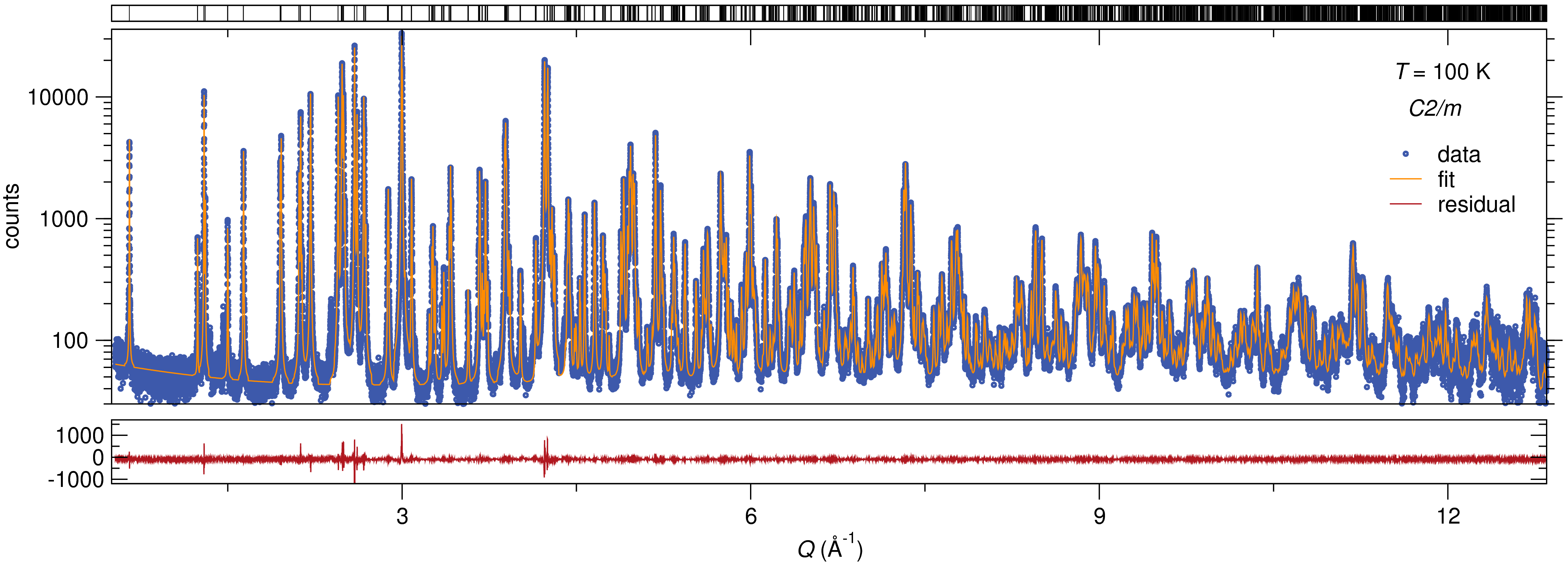}\\
\caption{(Color online) Powder synchrotron X-ray diffraction and Rietveld refinement for Co$_{10}$Ge$_3$O$_{16}$ at $T$ = 100\,K using a monoclinic $C$2/$m$ model. The low-temperature monoclinic cell was determined through pattern indexing and its atom positions were derived from the high-temperature $R\bar{3}m$ model using group-subgroup theory. The small residual and good refinement figures of merit support the validity of the model.}
\label{fig:Rietveld}
\end{figure*}

\begin{figure}
\centering
\includegraphics[width=2.5in]{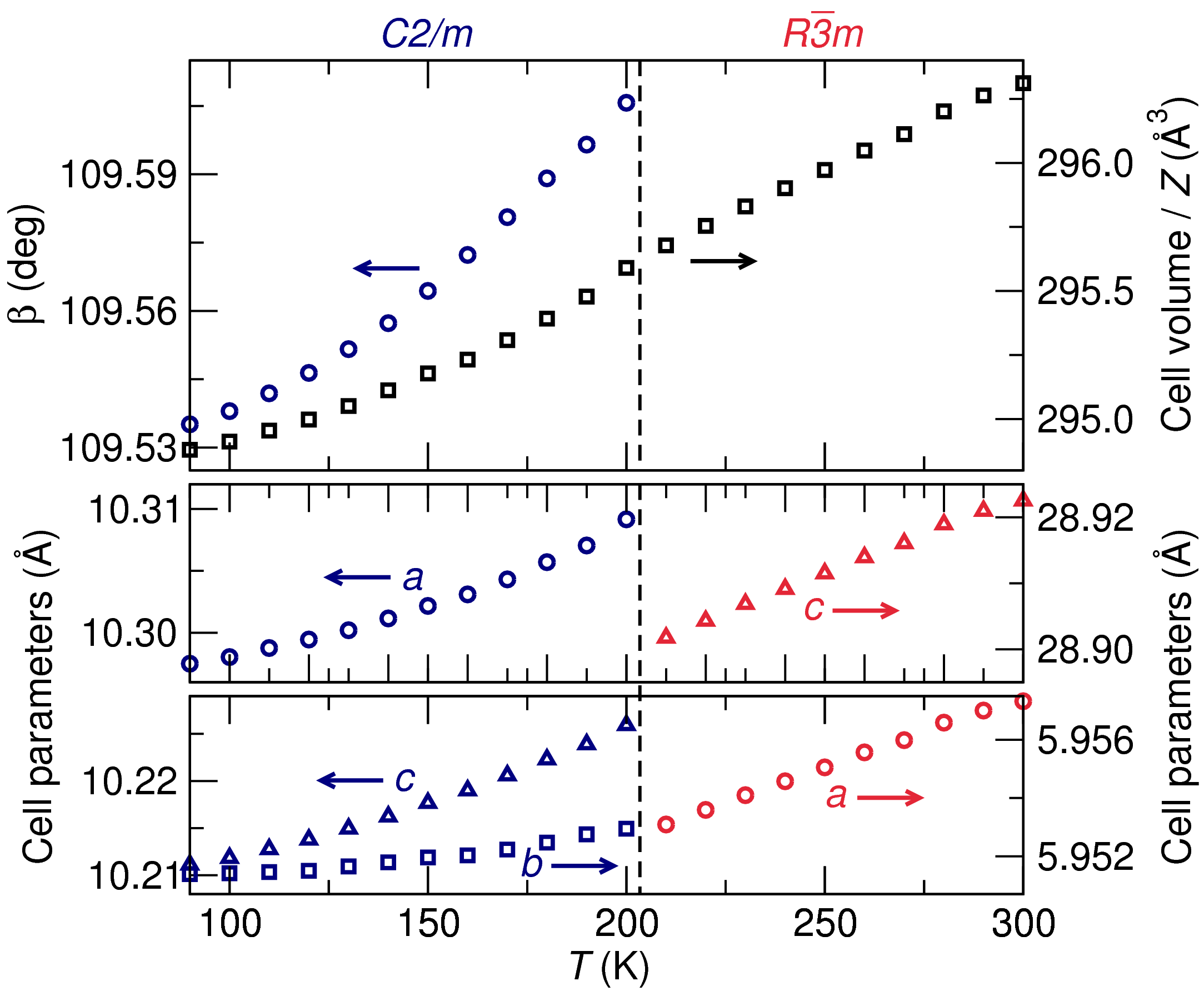}\\
\caption{(Color online) Structural parameters for Co$_{10}$Ge$_3$O$_{16}$ as a function of temperature, as determined by Rietveld refinement of powder synchrotron X-ray diffraction data. The dashed line indicates the phase boundary between the high-temperature $R\bar{3}m$ and low-temperature $C$2/$m$ structures. The $R\bar{3}m$ and $C$2/$m$ parameters are shown by the red and navy symbols, respectively, while the volume is displayed as the black squares. The arrows indicate the labeling axes.}
\label{fig:struct_param}
\end{figure}

Variable-temperature powder synchrotron X-ray diffraction was used to investigate the crystal structure of Co$_{10}$Ge$_3$O$_{16}$. The known crystal structure was confirmed by Rietveld refinement of the room-temperature diffraction pattern (not shown). The extracted room-temperature $R\bar{3}m$ structural parameters are $a$ = 5.957(2) and $c$ = 28.92(2), which agree well with the single-crystal study by Barbier.\cite{Barbier_AC95} The presence of a small 1.0 mol\,\% impurity of the GeCo$_2$O$_4$ phase was determined by our analysis. An unidentified and even smaller impurity phase was observed in the synchrotron X-ray data with peaks at $Q$ = 1.34(4), 1.42(8), 1.45(7), 1.46(7), 2.38(8), and 3.10(9) \AA$^{-1}$. We rule out any significant impurity effect on physical properties based on their small phase fractions as well as experiments on a second sample that exhibits the same behaviors yet does not contain any observable impurities. A structural phase transition was observed in the temperature evolution of the powder synchrotron X-ray diffraction. The onset of the transition is indicated by the splitting of the $R\bar{3}m$ (208) diffraction peak as temperature is lowered (Figure~\ref{fig:peak_evolution}). The diffraction data were fit by Rietveld refinement using the known $R\bar{3}m$ model above the transition and a new $C$2/$m$ model below. Unit cell parameters for the initial $C$2/$m$ model were determined by indexing the diffraction pattern. The $C$2/$m$ atom positions were derived from the $R\bar{3}m$ model using group-subgroup theory. A refinement using the $C$2/$m$ model at $T$ = 100\,K gives a good qualitative fit to the data, matching the observed peaks in both position and intensity (Figure~\ref{fig:Rietveld}). The small residual and $R_{\rm{Bragg}}$ support the validity of both the new low-temperature $C$2/$m$ model and the known high-temperature $R\bar{3}m$ model. The extracted structural parameters for $C$2/$m$ at 100\,K are listed in Table~\ref{tab:C2/m}. The structural transition involves a slight elongation of the CoO$_6$ octahedra along the apical axis, breaking the rhombohedral symmetry. Bond valence sum calculations indicate the ion valences expected from the chemical formula, namely Co$^{2+}$, Ge$^{4+}$, and O$^{2-}$, for both structures. The structural transition was determined to occur between $T$ = 200 and 210\,K by comparing fits of the diffraction data using the high- and low-temperature structural models. This temperature is consistent with the $T_{\rm{N}}$ = 203\,K from the magnetometry experiments discussed later. The unit cell parameters decrease smoothly with temperature, as expected for a typical positive coefficient of thermal expansion (Figure~\ref{fig:struct_param}). A change of slope, but not a discontinuity, is observed at $T_{\rm{N}}$ in the plot of cell volume as a function of temperature; this is typical of a second-order phase transition.

%Which symmetry element is broken?

%Any changes in Ge tetrahedra or octahedra?

%Cannot be ferroelectric because C2\m is centrosymmetric

%The ionic radius of Co2+ in Co10Ge3O16 is consistent with high spin.

%Udod lattice parameters are exactly the same as Barbier's...

% 1.8 mol% from XND or 0.9 mol% from GSAS GeCo2O4, and tiny peaks that are unaccounted for - 1.34421 (5.07), 1.4278 (5.38), 1.45745 (5.493), 1.46723 (5.529), 2.39 (9.00), 3.109 (11.73)

\begin{table}
\caption{Structural parameters of Co$_{10}$Ge$_3$O$_{16}$ at $T$~=~100\,K, as determined by Rietveld refinement of powder synchrotron X-ray diffraction data. Space group: $C$2/$m$, $a$ = 10.3021(0) \AA, $b$ = 5.9537(4) \AA, $c$ = 10.2158(1) \AA, and $\beta$ = 109.5(4) $^{\circ}$. Figures of merit: $\chi^2$ = 1.93, $R_{\rm{wp}}$ = 6.40\,\%, $R_{\rm{p}}$ = 5.07\,\%.}
\label{tab:C2/m}
\centering
\begin{tabular}{llllllllllllll}\\
\hline
\hline
Site &$x$ &$y$ &$z$ &$U_{\rm{iso}}$ (\AA$^2$) \\
\hline
Ge1 &0 &0 &0 &0.001(1)\ \\
Ge2 &0.1892(9) &0 &0.5679(4) &0.001(5)\ \\
Co1 &0 &0.5 &0.5 &0.004(0)\ \\
Co2 &0.25 &0.25 &0 &0.001(7)\ \\
Co3 &0 &0.5 &0 &0.002(5)\ \\
Co4 &0.0039(3) &0.2451(0) &0.2562(9) &0.001(1)\ \\
Co5 &0.2483(3) &0 &0.2558(3) &0.001(1)\ \\
O1 &0.1266(8) &0 &0.3821(8) &0.003(7)\ \\
O2 &0.3766(2) &0 &0.1283(1) &0.004(7)\ \\
O3 &0.6136(7) &0.2761(1) &0.1146(2) &0.003(2)\ \\
O4 &0.8863(1) &0 &0.1118(8) &0.002(0)\ \\
O5 &0.3728(3) &0.2548(7) &0.3741(0) &0.003(9)\ \\
O6 &0.6293(9) &0 &0.3775(9) &0.005(0)\ \\

\hline
\hline
\end{tabular}
\end{table}

\subsection{Magnetism}

\begin{figure}
\centering
\includegraphics[width=2.5in]{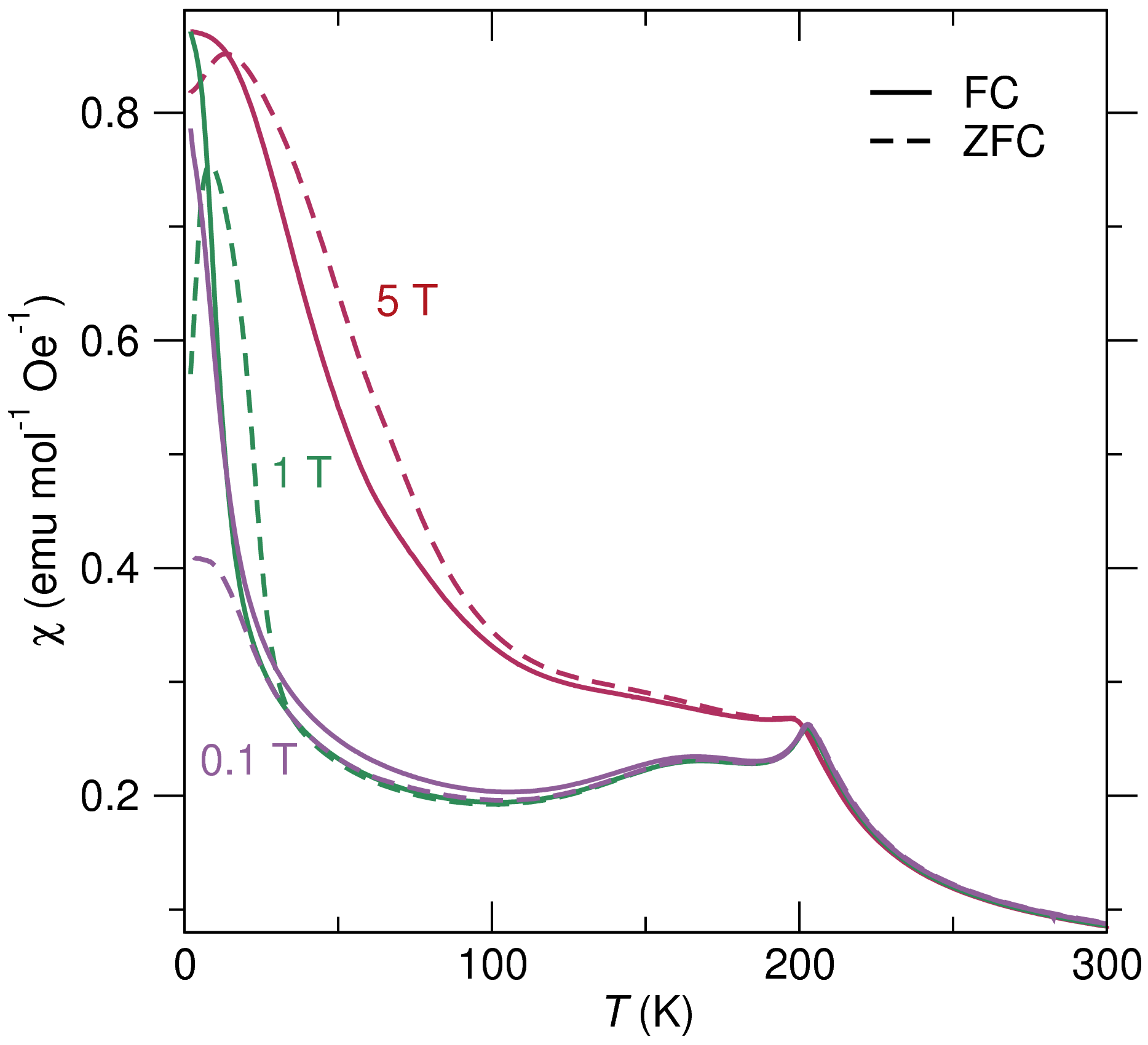}\\
\caption{(Color online) Magnetic susceptibility as a function of temperature for Co$_{10}$Ge$_3$O$_{16}$. The data were collected under $H$ = 0.1, 1, and 5\,T using a ZFC-FC procedure.}
\label{fig:susceptibility}
\end{figure}

\begin{figure*}
\centering
\includegraphics[width=6in]{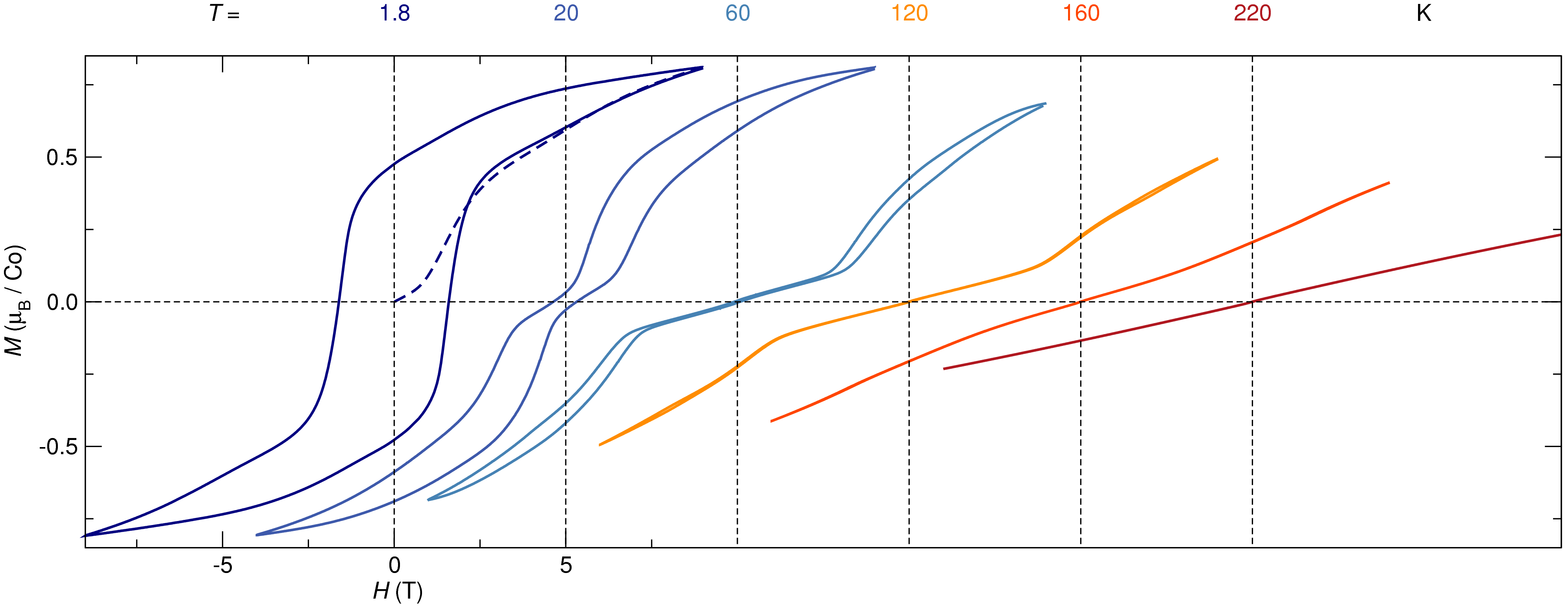}\\
\caption{(Color online) Magnetization versus field sweeps for Co$_{10}$Ge$_3$O$_{16}$ at different temperatures. The data for each temperature are offset by 5\,T increments along the $x$-axis to facilitate visualization. The virgin curve for the $T$~=~1.8\,K trace is dashed to distinguish it from the loop. Dashed black lines are provided to establish the graph origin for each data set.}
\label{fig:magnetization}
\end{figure*}

\begin{figure}
\centering
\includegraphics[width=2.5in]{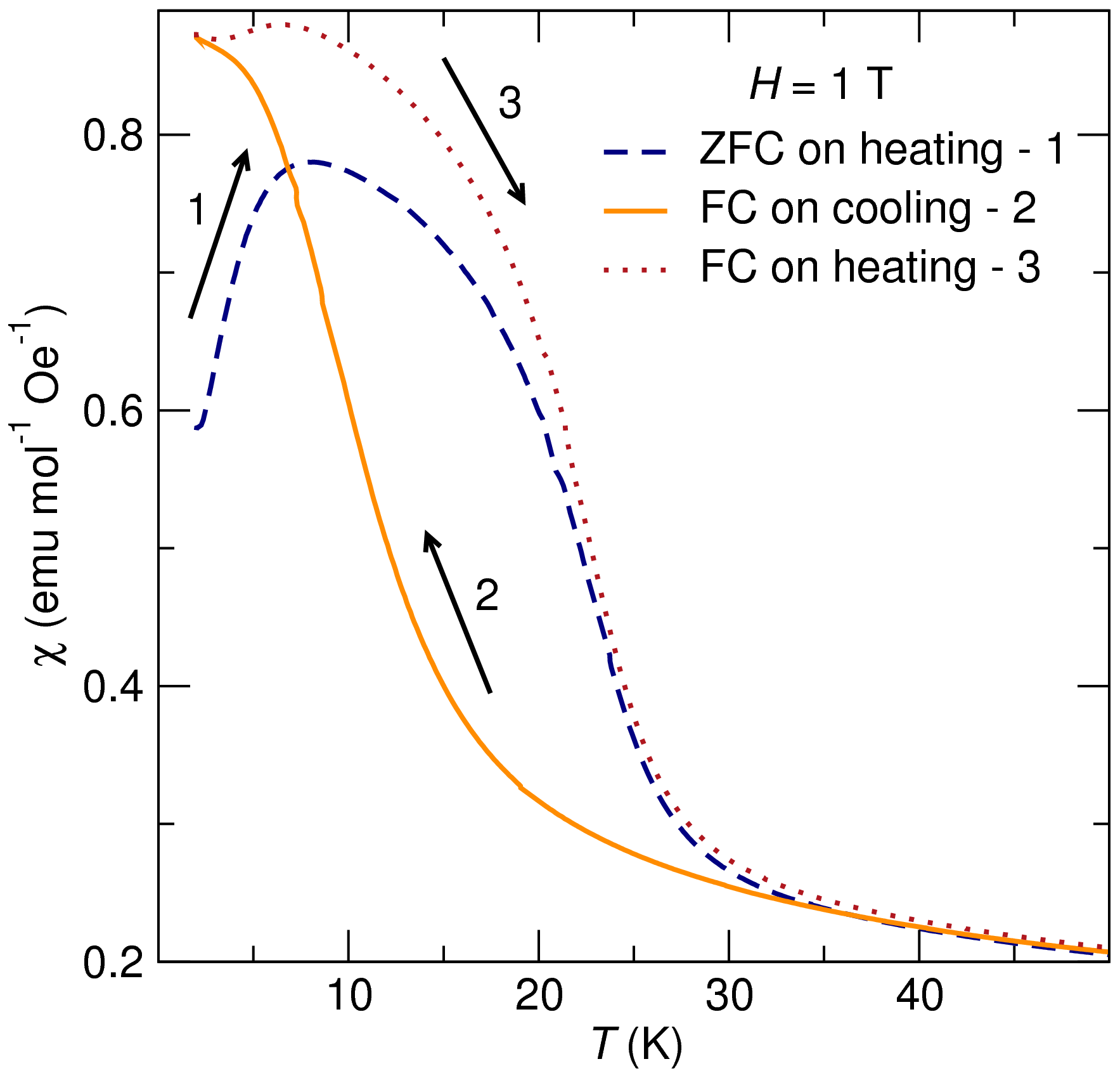}\\
\caption{(Color online) A field-induced first-order phase transition at low temperature in Co$_{10}$Ge$_3$O$_{16}$. The zero-field cooled (ZFC) trace is greater than the field-cooled (FC) for 7\,K $< T <$ 30\,K. This is a result of thermal hysteresis between the FC data collected on heating and cooling, with the curves offset by 11\,K at the largest point.}
\label{fig:first-order}
\end{figure}

The magnetic behavior of Co$_{10}$Ge$_3$O$_{16}$ involves phase transitions with both temperature and magnetic field. Magnetometry data, including both temperature ($\chi$-$T$, Figure~\ref{fig:susceptibility}) and field ($M$-$H$, Figure~\ref{fig:magnetization}) dependences, were collected from $T$ = 1.8\,K to 380\,K under applied magnetic fields up to $H$~=~9\,T. A standard zero-field cooled (ZFC) and field-cooled (FC) protocol was used for the temperature-dependent measurements. A Curie-Weiss fit of the magnetic susceptibility collected under $H$ = 1000\,Oe in the paramagnetic regime from $T$~=~300\,K to 380\,K gives $\mu_{\rm{eff}}$ = 4.26 $\mu_{\rm{B}}$/Co, assuming $g$ = 2, and $\Theta_{\rm{CW}}$ = +39.3\,K. The magnitude of $\mu_{\rm{eff}}$ is consistent with high-spin Co$^{2+}$, lying between its $S$ and $L+S$ values of 3.87 and 5.20. The positive $\Theta_{\rm{CW}}$ implies the presence of ferromagnetic interactions. However, as is known for GeCo$_2$O$_4$,\cite{Lashley_PRB08} the Curie-Weiss law does not strictly apply for octahedral Co$^{2+}$ because of low-lying excited states that affect the susceptibility. Consequently, the sign of $\Theta_{\rm{CW}}$ could be misleading. In low fields of $H$~=~100 (not shown) and 1000 Oe, a cusp in the $\chi$-$T$ occurs at $T_{\rm{N}}$~=~203\,K, suggesting that long-range antiferromagnetic order is established. Below this $T_{\rm{N}}$, a small broad hump, centered at $T$~=~170\,K, and a slight ZFC-FC splitting are observed. $\chi$ reaches a local minimum at $T$~=~100\,K and then increases as temperature is further lowered, with a sharp upturn at $T$~=~20\,K. Considering only the FC trace, this increase in $\chi$ might be interpreted as a Curie tail from a paramagnetic impurity. However, we also observe that the ZFC-FC splitting increases significantly below $T$~=~20\,K and that the ZFC exhibits a peak at $T$~=~9\,K. These features in the ZFC-FC indicate the onset of another phase transition involving uncompensated magnetism. As seen in the $H$ = 1 and 5\,T susceptibility curves, this low-temperature transition is strongly affected by the application of a large magnetic field, the details of which are discussed in the next paragraph. Isothermal $M$-$H$ sweeps are shown in Figure~\ref{fig:magnetization} for different temperatures down to $T$~=~1.8\,K. The curves are separated into three groups by temperature region and plotted in sequence to simulate the evolution with temperature. The $M$-$H$ is linear above the transition at $T$~=~220\,K, consistent with paramagnetism at low $B$/$T$. As temperature is decreased below $T_{\rm{N}}$ = 203\,K, the $M$-$H$ behavior remains linear with the slope following that expected from the $\chi$-$T$. However, a broad upturn in the $M$-$H$ at high field begins to evolve as $T$ passes beneath 180\,K. By $T$~=~120\,K, this upturn develops into a sharp kink at a critical field $H_{\rm{C}}$ = 3.9\,T. This phase transition involves hysteresis at high fields while the dependence remains linear and reversible at low fields. A ``butterfly'' shape is thus formed, with distinct loops existing at both high positive and negative fields. $H_{\rm{C}}$ decreases as temperature is further lowered and the hysteresis loops eventually merge starting at $T$~=~20\,K. The $M$-$H$ sweep at 1.8\,K exhibits a single hysteresis loop with a sizable coercivity of nearly 1.5\,T. Interestingly, the virgin magnetizing curve taken after ZFC lies outside the hysteresis loop. This unusual $M$-$H$ feature has previously been associated with irreversible domain wall motion in spinel oxides.\cite{Joy_JMMM00} Looking back at the $\chi$-$T$, we see that the measurement under $H$~=~5\,T exhibits a change in slope at $T_{\rm{N}}$, as opposed to a cusp, and that $\chi$ has a larger magnitude than in small fields, consistent with $H_{\rm{C}} <$ 5\,T.

%Field-induced transition becomes hysteretic with field below 

%Show dM/dH-H-T contour plot?
%Effect of heating/cooling rate on M-T?

%Try a ZFC-FC measurement where you only cool to ~100 K. Will there still be an offset?

%dChi/dT. Interesting dependence.

%What would it be if it was intermediate spin S=1/2?

The application of a large magnetic field generates a first-order phase transition in Co$_{10}$Ge$_3$O$_{16}$ at temperatures below 120\,K. This corresponds to the same temperature at which the field-induced transition in the $M$-$H$ becomes sharp and hysteretic. The first-order nature of the transition was first suspected because of an unusual feature seen in the high-field susceptibility data where the ZFC trace is larger in magnitude than the FC (Figure~\ref{fig:susceptibility}). For measurement under $H$ = 1 this ZFC $>$ FC for a region 7\,K $< T <$ 30\,K, while it is 14\,K $< T <$ 120\,K for $H$ = 5\,T. These data were collected using a normal ZFC-FC procedure where the ZFC is collected on warming and the FC is collected on cooling, however following this unusual ZFC $>$ FC observation, we measured a second FC curve on heating to investigate any history effect. A thermal hysteresis, indicating a latent heat, is observed between the FC data collected on heating and cooling, with the curves offset in the $H$ = 1\,T measurement by 11\,K at the largest point (Figure~\ref{fig:first-order}). This latent heat is only observed for measurements with $H$ $>$ 0.3\,T. For $H$ = 1\,T, the bifurcation point between the ZFC and FC curves measured on heating occurs at $T$ = 22\,K. The first-order nature of this magnetic transition necessitates the presence of a structural component. Such a strong first-order phase transition is uncommon in oxides containing Co(II).

%Similar first-order transitions?

\begin{figure}
\centering
\includegraphics[width=2.5in]{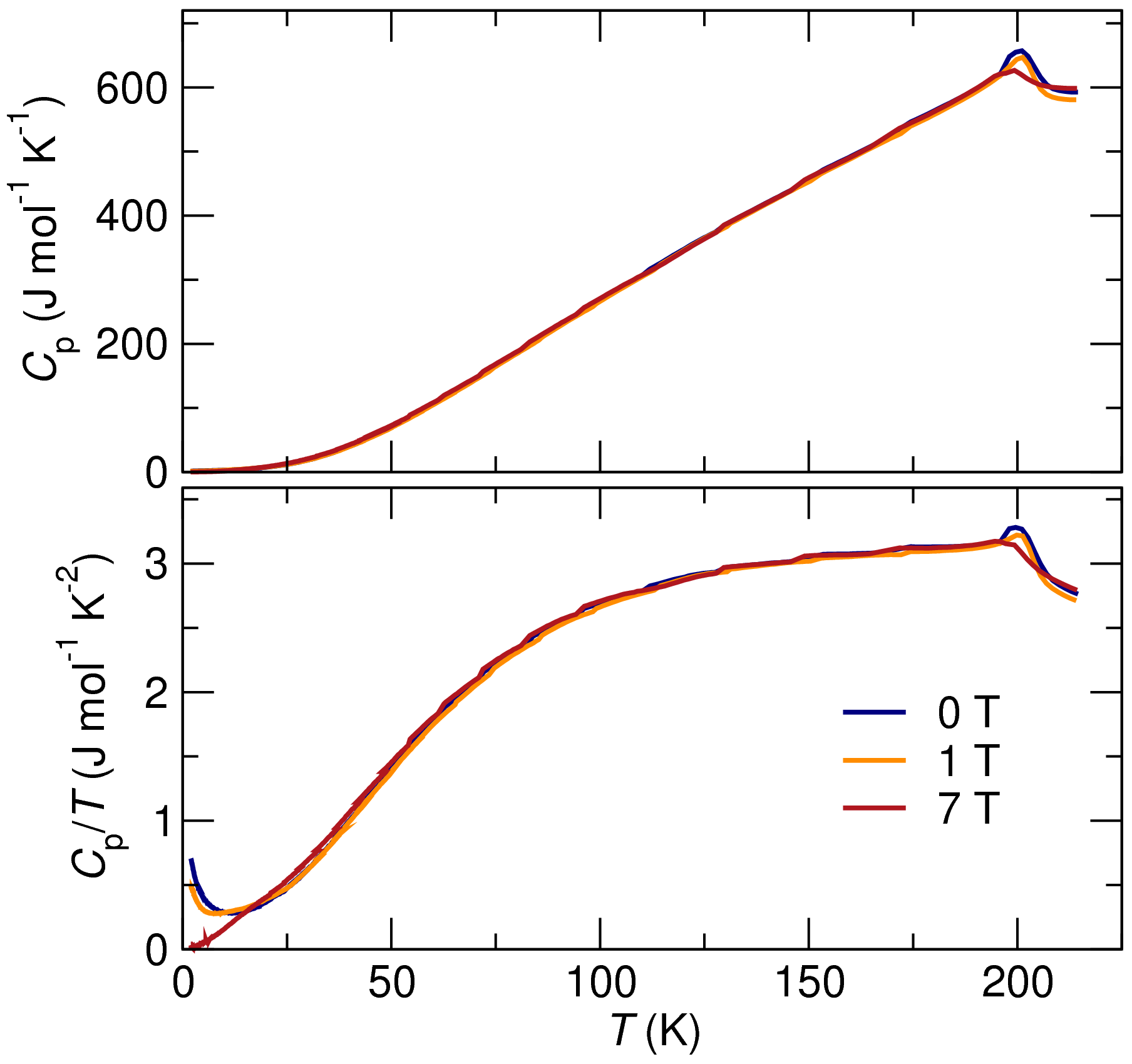}\\
\caption{(Color online) Heat capacity for Co$_{10}$Ge$_3$O$_{16}$. The top panel displays the heat capacity collected under different magnetic fields, while the bottom shows the heat capacity normalized by temperature. An anomaly is observed at $T$ = 203\,K, corresponding to the magnetostructural phase transition. This transition temperature is lowered to $T$ = 199\,K with the application of $H$ = 7\,T. A small upturn in $C_{\rm{p}}$/$T$ is seen below 15\,K and is suppressed by magnetic field.}
\label{fig:heat_capacity}
\end{figure}

\begin{figure*}
\centering
\includegraphics[width=6in]{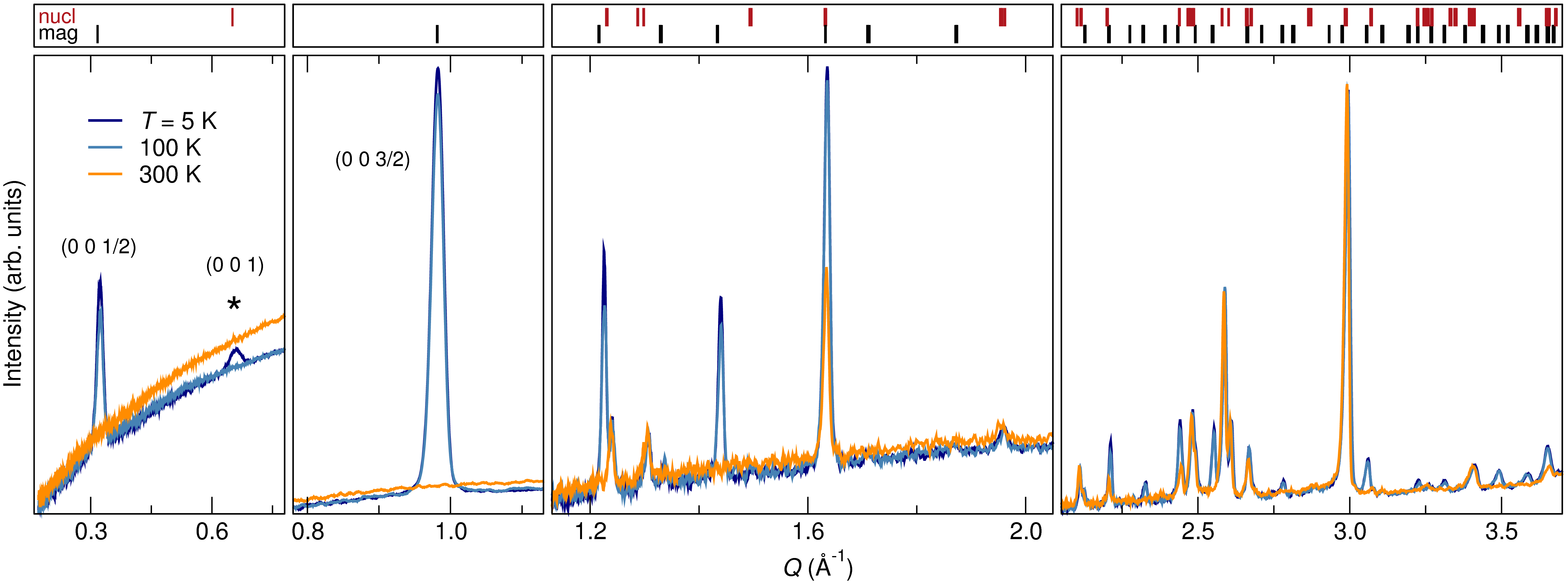}\\
\caption{(Color online) Powder neutron diffraction for Co$_{10}$Ge$_3$O$_{16}$ at different temperatures. Magnetic Bragg reflections of $k_{\rm{M}}$ = (00$\frac{1}{2}$) emerge beneath $T_{\rm{N}}$ = 203\,K. The peak positions of the nuclear and $k_{\rm{M}}$ = (00$\frac{1}{2}$) magnetic phases are shown in the panel above the patterns, colored red and black respectively. The $T$ = 5 and 100\,K patterns are quite similar, however a close look reveals a new magnetic peak at $Q$ = 0.65 \AA$^{-1}$, the (001) nuclear position, as well as additional magnetic intensity at other nuclear reflections, which is consistent with the onset of an additional $k_{\rm{M}}$ = (000). This development occurs for $T <$ 20\,K and is associated with the uncompensated magnetism that is observed concomitantly. Each panel displays data from a separate detector bank with different resolution.}
\label{fig:neutron}
\end{figure*}

Heat capacity measurements were used as a complementary method to investigate the phase transitions (Figure~\ref{fig:heat_capacity}). An anomaly is observed at $T_{\rm{N}}$ = 203\,K under zero applied field, consistent with the magnetostructural transition seen in the diffraction and magnetometry data. The field dependence of the high-temperature measurements reveals that the transition temperature is decreased by a few Kelvin upon the application of $H$ = 7\,T, consistent with antiferromagnetic theory. No obvious features are seen down to $T$ = 2\,K in the zero-field trace, nor in the measurements under $H$ = 1 or 7\,T, despite the presence of the field-induced first-order transition revealed by magnetometry. However, the $C_{\rm{p}}$/$T$ shows an upturn below 15\,K that is suppressed with field. It is difficult to assign this second anomaly to a specific magnetic, structural, or electronic origin. Its weak nature suggests that the low-temperature magnetic transition is a spin reorientation and thus does not involve a change in entropy. The entropy loss could instead be a result of a structural transition; a subtle change in octahedral shape, which is already distorted beneath $T_N$, as opposed to an order-disorder transition for example, would generate only a small response. It is also possible that this measurement is probing the Kramers' doublet electronic ground state of octahedral Co(II). Finally, the first-order nature is problematic for the thermal relaxation mechanism of our heat capacity measurement. We are thus planning to investigate the heat capacity with a technique that directly measures heat flux.

% What about isothermal Cp-H? Would it be similar to the M-H with a significant upturn and hysteresis?

Powder neutron diffraction (PND) was conducted to further elucidate the magnetism of Co$_{10}$Ge$_3$O$_{16}$ (Figure~\ref{fig:neutron}). We do not report the magnetic structure here, however we discuss the implications of some important observations. Magnetic Bragg reflections materialize beneath $T_{\rm{N}}$ = 203\,K and confirm the onset of antiferromagnetism. The structural component of the phase transition known from the synchrotron X-ray experiments is just barely evident as a subtle peak broadening in the neutron data. The peak positions index to a magnetic $k$-vector $k_{\rm{M}}$ = (00$\frac{1}{2}$), indicating that the magnetic unit cell is commensurate with the low-temperature monoclinic nuclear structure but that it is doubled along the $c$-axis. The intensities and positions of the peaks evolve smoothly with temperature down to $T$ = 20\,K at which point a small new peak emerges at $Q$ = 0.65\,\AA$^{-1}$, along with some subtle modulations of intensity in other reflections. We attribute this new reflection to a magnetic origin because although its $Q$ value corresponds to the (001) plane of the nuclear unit cell, and thus it could be a nuclear peak, no intensity is observed above background at this location for $T >$ 20\,K. A careful comparison of the $T$ = 5 and 100\,K patterns reveals that magnetic intensity is also added to other solely nuclear reflections such as $Q$ = 1.24, 1.96, 3.41\,\AA$^{-1}$. Additional support for a magnetic origin is seen in the low-field magnetometry data, where a FC increase, ZFC-FC separation, and ZFC peak are observed with the same onset temperature as this new magnetic peak. The occurrence of this magnetic Bragg reflection at the nuclear (001) suggests that it belongs to $k_{\rm{M}}$ = (000). Given the uncompensated ferri- or ferromagnetic nature of this $k_{\rm{M}}$ = (000) seen in the low-field magnetometry data, we suggest that it is connected to the transition occurring at large magnetic field. We thus hypothesize that the application of a magnetic field will enhance $k_{\rm{M}}$ = (000) at the expense of $k_M$ = (00$\frac{1}{2}$).

The spontaneous, and likely field-induced, $k_M$ = (000) uncompensated magnetism of Co$_{10}$Ge$_3$O$_{16}$ appears to be ferrimagnetic in nature. We observe that $M$ = 0.82 $\mu_{\rm{B}}$/Co at $T$ = 1.8\,K and $H$ = 9\,T, and that the $M$-$H$ is trending toward saturation at an intermediate $M_{\rm{S}}$ = 1.0 $\mu_{\rm{B}}$/Co, one third of the expected value for $S$ = 3/2 Co(II). This value of $M_{\rm{S}}$ suggests a ferrimagnetic state where partial compensation of the moment arises from the magnetic structure. An alternative explanation to explain the reduced $M_{\rm{S}}$, however, is that Co$_{10}$Ge$_3$O$_{16}$ adopts a ferromagnetic low-spin $S$ = 1/2 state. While low-spin Co(II) is found in CoS$_2$ and CoSe$_2$ where the crystal field is larger,\cite{Adachi_JPSJ69} a spin transition is unheard of in extended, non-molecular Co(II) systems and would involve significant changes in crystal structure. Additionally, given the large magnetic field needed to drive a spin-state transition in the related Co(III) compound LaCoO$_3$,\cite{Altarawneh_PRL12} a similar transition seems unlikely to occur under our experimental conditions. The spontaneous emergence of this ferrimagnetism and its coexistence with the antiferromagnetism is unusual for Co(II)-containing oxides and even for inorganic compounds in general.
%However, Co(II) is not Co(III)
%The existence of a spin-state transition upon application of a magnetic field could be investigated by analyzing electron paramagnetic resonance spectra, by examining the refined magnetic moment from neutron diffraction, or by looking for structural changes in X-ray diffraction.

\begin{figure}
\centering
\includegraphics[width=2.5in]{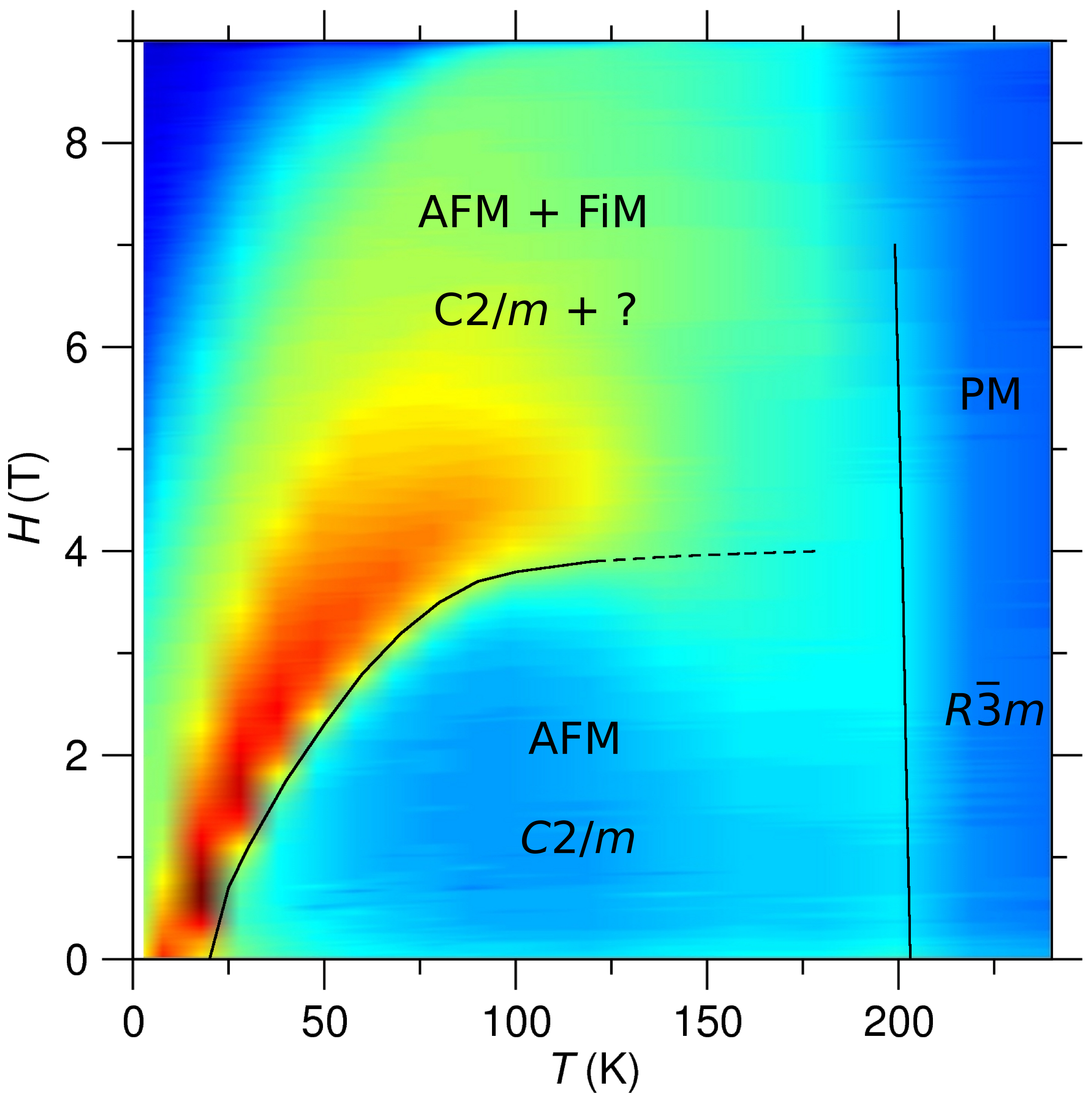}\\
\caption{(Color online) Structural and magnetic $H$-$T$ phase diagram for Co$_{10}$Ge$_3$O$_{16}$. The contour plot of log(d$M$/d$H$), taken from the first quadrant of the $M$-$H$ plot, demonstrates the development of ferrimagnetism (FiM). Magnetometry and neutron diffraction experiments revealed the spontaneous emergence of FiM for $T <$ 20\,K. The phase boundary between antiferromagnetism (AFM) and paramagnetism (PM) was observed in both the magnetic and heat capacity data. The indicated crystal structure space groups were solved by refinement of powder synchrotron X-ray diffraction. The FiM phase must have a different nuclear structure than the AFM phase because the transition between them is first-order. The solid and dashed lines indicate first- and second-order nature, respectively.}
\label{fig:phase_diagram}
\end{figure}

The structural and magnetic $H$-$T$ phase diagram of Co$_{10}$Ge$_3$O$_{16}$, as determined by analysis of variable-temperature powder synchrotron X-ray and neutron diffraction, magnetometry, and heat capacity experiments, is presented in Figure~\ref{fig:phase_diagram}. The free energy landscape gives rise to an $H_{\rm{C}}$ that decreases with decreasing temperature, which is opposite to that observed in typical antiferromagnetic systems such as CoCl$_{2}\cdot$2H$_2$O, the archetypal Co(II) metamagnet.\cite{Kobayashi_JPSJ64} This difference likely arises due to the spontaneous tendency of Co$_{10}$Ge$_3$O$_{16}$ towards a ferrimagnetic ground state. The behavior of Co$_{10}$Ge$_3$O$_{16}$ is also strikingly dissimilar to other compounds in the Co-Ge-O system, namely spinel GeCo$_2$O$_4$ and pyroxene CoGeO$_3$,\cite{Redhammer_PCM10} which are both antiferromagnets that exhibit a reversible metamagnetic transition to a ferromagnetic state. Additionally, the $T_N$ = 203\,K of Co$_{10}$Ge$_3$O$_{16}$ is much larger than the 20\,K and 35\,K of GeCo$_2$O$_4$ and CoGeO$_3$, respectively. While Co$_{10}$Ge$_3$O$_{16}$ contains octahedral Ge$^{4+}$ that is not present in the other Co-Ge-O compounds, the contrasting behavior is likely due to the higher density of Co and its different connectivity. Despite these distinct behaviors, a region of phase coexistence between antiferromagnetism and ferrimagnetism exists in Co$_{10}$Ge$_3$O$_{16}$ that is similar to a region of antiferromagnetism and ferromagnetism in GeCo$_2$O$_4$.\cite{Matsuda_JPSJ11} One might ascribe this broad two-phase region to powder averaging, especially when comparing to CoCl$_{2}\cdot$2H$_2$O,\cite{Kobayashi_JPSJ64} for example, which exhibits sharp transitions in its $M$-$H$. Transitions from AFM to FiM and from FiM to FM occur at $H$ = 0.32 and 0.46\,T, respectively, for measurements with $H \parallel b$, but no transitions if the field is along the $a$ or $c$ axes. However, single-crystal measurements on the related GeCo$_2$O$_4$ show that it does not exhibit significantly anisotropic magnetism despite the single-ion anisotropy of Co(II).\cite{Hoshi_JMMM07} It is instead likely that the magnetic anisotropy is intimately related to the magnetic structure, and thus the large phase space of Co$_{10}$Ge$_3$O$_{16}$ and GeCo$_2$O$_4$ may be a consequence of their high density of magnetic ions and the presence of many exchange pathways.

\section{Conclusions}

The crystal structure of Co$_{10}$Ge$_3$O$_{16}$ is an intergrowth with alternating layers of spinel and rock salt. A magnetostructural transition occurs at $T_{\rm{N}}$ = 203\,K, evidenced by magnetometry, heat capacity, and powder synchrotron X-ray experiments. This transition gives rise to long-range antiferromagnetic order with a rhombohedral-to-monoclinic symmetry change resulting from a slight distortion of the CoO$_6$ octahedra. The antiferromagnetism is characterized by $k_{\rm{M}}$ = (00$\frac{1}{2}$), as revealed by powder neutron diffraction, and below $T$ = 20\,K a small uncompensated component with $k_{\rm{M}}$ = (000) emerges spontaneously and coexists with the antiferromagnetism. The Ising-like spins of Co$^{2+}$ reorient themselves below $T_{\rm{N}}$ upon the application of a large magnetic field. This transition occurs at $H_{\rm{C}}$ = 3.9\,T for $T$ = 180\,K and gives rise to a broad upturn in the $M$-$H$, while at $T$ = 120\,K it sharpens into a kink that exhibits hysteresis. A ``butterfly'' loop is thus formed, with linear and reversible behavior at low fields and hysteresis loops at high fields. The loops at positive and negative fields in the $M$-$H$ merge beneath $T$ = 20\,K because $H_{\rm{C}}$ decreases as temperature decreases. A latent heat is observed in temperature-dependent measurements and indicates that this metamagnetic transition is strongly first-order. The low-temperature $M$-$H$ trends toward saturation at $M_{\rm{S}}$ = 1.0 $\mu_{\rm{B}}$/Co even though this is only one third the value expected for high-spin Co(II). This reduced $M_{\rm{S}}$ points toward the ferrimagnetic nature of the field-induced state rather than a spin-state transition to a $S$ = 1/2 ferromagnet. The structural and magnetic $H$-$T$ phase diagram is not typical of inorganic Co(II) compounds nor of other transition metal oxides.

\section{Acknowledgments}
PTB is supported by the NSF Graduate Research Fellowship Program. RS and PTB acknowledge the support of the NSF through DMR 1105301. We acknowledge the use of MRL Central Facilities which are supported by the MRSEC Program of the NSF under Award No. DMR 1121053; a member of the NSF-funded Materials Research Facilities Network (www.mrfn.org). Use of data from the 11-BM beamline at the Advanced Photon Source was supported by the U.S. Department of Energy, Office of Science, Office of Basic Energy Sciences, under Contract No. DE-AC02-06CH11357. This work has benefited from the use of HIPD at the Lujan Center at the Los Alamos Neutron Science Center, funded by the DOE Office of Basic Energy Sciences. Los Alamos National Laboratory is operated by Los Alamos National Security LLC under DOE Contract DE-AC52-06NA25396. 

%A portion of this work was performed at the National High Magnetic Field Laboratory, which is supported by the NSF Cooperative Agreement DMR-0654118, the State of Florida, and the U.S. DoE BES program “Science in 100 T”.

\bibliography{Co10Ge3O16}

\end{document}